\pdfoutput=1
\documentclass[11pt,twoside,a4paper,cmspaper,final,collab]{cms-tdr}
\def\svnVersion{79749c2}\def\svnDate{2026/01/12}\def\cmsCernNoTag{CERN-EP-2025-051}\def\cmsCernDate{\today}\def\cmsMessage{Published in Physical Review Letters as \href{http://dx.doi.org/10.1103/2ssw-wwyy}{\doi{10.1103/2ssw-wwyy}.}}
\begin{document}\cmsNoteHeader{HIN-24-009}

\newcommand {\KK} {\ensuremath{\PKp \PKm}\xspace}
\newcommand {\pipi} {\ensuremath{\PGpp \PGpm}\xspace}
\newcommand {\mKK} {\ensuremath{m_{\KK}}\xspace}
\newcommand {\Ncoh} {\ensuremath{N^{\text{raw}}_{\PGf}}\xspace}
\newcommand {\Sphi} {\ensuremath{\mathrm{S_{\PGf}}}\xspace}
\newcommand {\Sjpsi}{\ensuremath{\mathrm{S^{\PJGy}}}\xspace}
\newcommand {\PbPb} {\ensuremath{\text{PbPb}}\xspace}
\newcommand {\pb}   {\ensuremath{\text{Pb}}\xspace}
\newcommand {\fI} {\ensuremath{f_{\text{I}}}\xspace}
\newcommand {\DSigDyPhi}  {\dd{\sigma^{\text{coh}}_{\PGf}}{y}\xspace }
\newcommand {\STARlight}{\textsc{STARlight}\xspace}
\newcommand {\dEdx}{\ddinline{E}{x}\xspace}

\cmsNoteHeader{HIN-24-009}

\title{Observation of coherent \texorpdfstring{\Pgf}{phi} meson photoproduction in ultraperipheral \PbPb collisions at \texorpdfstring{$\sqrtsNN = 5.36\TeV$}{sqrt(sNN) = 5.36 TeV}
}

\date{\today}

\abstract{
The first observation of coherent \Pgf meson photoproduction off heavy nuclei is presented using ultraperipheral lead-lead collisions at a center-of-mass energy per nucleon pair of 5.36\TeV. The data were collected by the CMS experiment and correspond to an integrated luminosity of 1.62\mubinv. The \Pgf meson signals are reconstructed via the \KK decay channel. The production cross section is presented as a function of the \Pgf meson rapidity in the range $0.3 < \abs{y} < 1.0$, probing gluons that carry a fraction of the nucleon momentum ($x$) around $10^{-4}$. The observed cross section exhibits little dependence on rapidity and is significantly suppressed, by a factor of ${\sim} 5$, compared to a baseline model that treats a nucleus as a collection of free nucleons.
Theoretical models that incorporate the nuclear shadowing effect generally provide a better description of the \Pgf data than those incorporating gluon saturation.
This study establishes a powerful new tool for exploring nuclear effects and nuclear gluonic structure in the small-$x$ regime at a unique energy scale bridging the perturbative and nonperturbative quantum chromodynamics domains.
}

\hypersetup{%
pdfauthor={CMS Collaboration},%
pdftitle={Observation of coherent phi meson photoproduction in ultraperipheral PbPb collisions at sqrt(s[NN]) = 5.36 TeV},%
pdfsubject={CMS},%
pdfkeywords={CMS, UPC, photoproduction, gluon saturation, nuclear shadowing}} % limit six total

\maketitle 
Understanding the fundamental structure of nuclear matter is a central goal of quantum chromodynamics (QCD). At high energies or small $x$ values, where $x$ represents the momentum fraction carried by a parton within a nucleon, 
gluon densities rise rapidly, leading to gluon dominance in the partonic structure~\cite{H1:2015ubc}. It is theorized that this growth of gluon density is primarily driven by gluon splitting; however, at small $x$, non-linear effects such as gluon recombination are predicted to counterbalance this increase, leading to the phenomenon of ``gluon saturation''~\cite{Iancu_2004,Gelis:2010nm}. Heavy nuclei, with significantly higher gluon densities than free nucleons, provide an ideal setting to investigate this phenomenon~\cite{Accardi:2012qut}. Furthermore, nuclear shadowing, which may arise from coherent multi-nucleon interactions, can also reduce the effective gluon density in nuclei compared to that in free nucleons, thereby also influencing small-$x$ physics~\cite{Frankfurt:2005mc}.
A combined treatment of gluon saturation and nuclear shadowing, though challenging, would be valuable for future theoretical developments.

In ultraperipheral collisions (UPCs) of high-energy heavy ions, a photon emitted by one nucleus can fluctuate into a virtual quark-antiquark pair and interact diffractively with the other nucleus, producing a vector meson (VM) such as a $\PGrzP{770}$, \Pgf, \PJGy, or \PGU meson~\cite{Baltz:2007kq, Adamczyk:2017vfu, Afanasiev:2009hy, ALICE:2014eof, Sirunyan:2018sav, Sirunyan:2019nog, STAR:2022wfe, Abelev:2012ba, Abbas:2013oua, Acharya:2019vlb, Acharya:2019vlb, ALICE:2021gpt, Khachatryan:2016qhq, LHCb:2021bfl, LHCb:2022ahs}. In what follows, the \Pgf meson will be referred to as the \PGf meson. This photoproduction process is sensitive to the nuclear gluon distribution~\cite{Guzey:2013jaa, Klein:2020fmr}.
If the photon interacts with the entire nucleus, the process is called coherent; if it interacts with its internal constituents, it is called incoherent.
At the leading order in perturbative QCD (pQCD), the cross section for coherent VM production scales as $\sigma_{\mathrm{VM}} \propto (x G(x, Q^2))^2$, where $G(x, Q^2)$ is the gluon density at an energy scale $Q^2$ that is often taken to be ${\sim} (Q_0^2 + m_{\mathrm{VM}}^2)$, with $Q_0 {\sim} 0\GeV$ for quasi-real photons~\cite{Klein:2008di,Accardi:2012qut,Jenkovszky:2023jrh}. The VM mass ($m_{\mathrm{VM}}$) is therefore the primary determinant of the energy scale in QCD calculations. The onset of gluon saturation and the applicability of pQCD are highly sensitive to this scale.

While the \PJGy meson probes gluon distributions at relatively large scales ($Q^2 \gtrsim 10\GeV^2$), the \PGf meson, composed of a strange quark-antiquark pair with a mass of 1.019\GeV~\cite{ParticleDataGroup:2024cfk}, probes an intermediate regime where both perturbative and nonperturbative QCD dynamics are relevant. 
Varying the vector-meson mass effectively scans different interaction scales, providing sensitivity to the transition between dilute and dense gluonic matter. 
Because gluon-saturation models feature an explicit saturation scale $Q_s(x,A)$, whereas nuclear-shadowing models do not, the \PGf meson offers a unique test of their scale dependence. 
Measuring the coherent photoproduction of \PGf mesons is challenging due to the very low transverse momentum of 
the decay kaons ($\pt^\text{trk} < 0.15\GeV$), which limits the detector acceptance and particle identification efficiency. 
This has prevented \PGf mesons from being used as probes in previous UPC studies.

This Letter presents the first measurement of coherent \PGf meson photoproduction in heavy ion UPCs.
The \PGf meson is reconstructed via the decay channel $\PGf \to \KK$ using \PbPb collision data collected by the CMS detector at the CERN LHC at a center-of-mass energy per nucleon pair of $\sqrtsNN = 5.36\TeV$.
The dataset corresponds to an integrated luminosity of $1.62 \pm 0.10 \mubinv$.
The differential cross section is reported as a function of \PGf meson rapidity for the range $0.3<\abs{y}<1.0$, probing $x$ values around $10^{-4}$.
Acceptance and efficiency limitations in reconstructing \PKpm mesons restrict the measurable coherent \PGf meson yield to this rapidity range.
The results are compared with a wide range of theoretical predictions that include gluon saturation or nuclear shadowing effects.
The tabulated results are provided in the HEPData record~\cite{hepdata.156183}.

The central feature of the CMS apparatus~\cite{CMS:2008xjf} is a superconducting solenoid with an internal diameter of 6\unit{m}, 
providing a magnetic field of 3.8\unit{T}. 
Within the solenoid volume are a silicon pixel and strip tracker~\cite{CMS:2014pgm}, a lead tungstate crystal electromagnetic calorimeter~\cite{CMS:2020uim}, 
and a brass and scintillator hadron calorimeter, each composed of a barrel and two endcap sections, providing pseudorapidity ($\eta$) coverage $\abs{\eta} < 3.0$.
The hadron forward (HF) calorimeter~\cite{Bayatian:2006jz} is divided into two halves, one on either side of the interaction region, 
with each half segmented in $0.175 \times 0.175$ ($\Delta\eta{\times}\Delta\phi$) ``towers.'' 
This detector extends the $\eta$ coverage provided by the barrel and endcap detectors into the range $3.0 < \abs{\eta} < 5.2$.
The CMS apparatus can be used to detect and identify electrons, muons, photons, and hadrons~\cite{Bayatian:2006jz,CMS:2020uim,CMS:2018rym,CMS:2014pgm}. 
The silicon pixel and strip tracker is capable of reconstructing charged particles with $\pt^\text{trk}$ values down to 0.05\GeV. 
A more detailed description of the CMS detector, 
together with a definition of the coordinate system used and the relevant kinematic variables, can be found in Refs.~\cite{CMS:2008xjf,CMS:2023gfb}.

Events are selected using a zero-bias trigger with an additional online requirement of at least one reconstructed pixel track, achieving about 90\% efficiency 
for detecting at least one daughter kaon from a \PGf meson in the analysis phase space.
In a subsequent offline selection, events must have a primary interaction vertex, formed by at least two tracks coming from the collision, 
that is within 25\cm along the beam axis and within 2\cm in the transverse plane from the detector center.

To select UPC events, the largest energy deposit in any of the HF calorimeter towers must fall below a tower specific threshold determined from empty bunch crossing events.
This eliminates nearly all hadronic events while maintaining negligible loss of UPC signal efficiency~\cite{CMS:2020skx}.
Furthermore, events are required to have exactly two ``high purity tracks'' (as defined in Ref.~\cite{CMS:2014pgm}) with opposite electric charges, and with $\pt^\text{trk} > 0.05\GeV$ and $\abs{\eta} < 2.4$.

\begin{figure}[hbt!]
   \centering
   \includegraphics[keepaspectratio,width=0.95\linewidth]{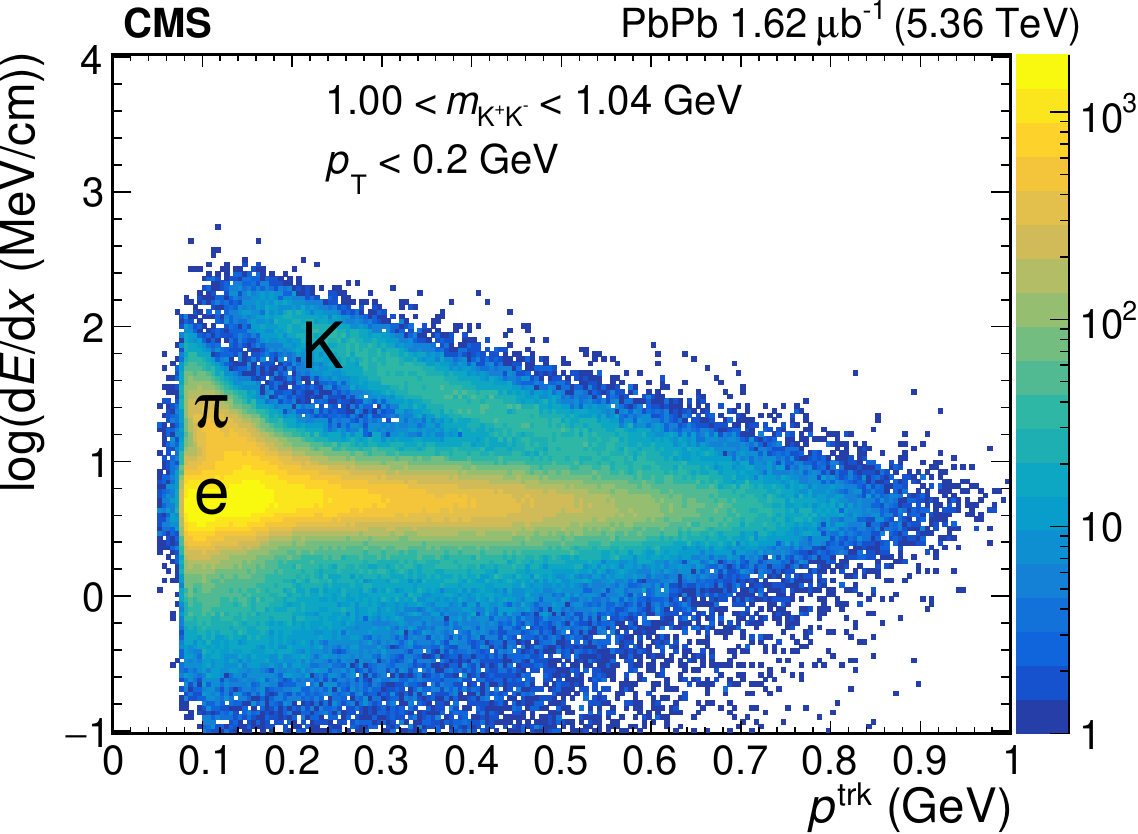}
   \caption{The log(\dEdx) distribution of daughter tracks of coherent \PGf candidates with $\pt{<}0.2\GeV$, selected within the signal invariant mass region, as a function of the daughter track total momentum, $p^\text{trk}$.}
   \label{fig:dedx}
\end{figure}

The kaon identification is performed using the energy loss information (\dEdx) measured by the pixel detector,  as illustrated in Fig.~\ref{fig:dedx},
following a probability-based method described in Refs.~\cite{CMS:2012xvn,TOTEM:2024aso}.
Here and throughout the rest of the paper, \pt refers to the transverse momentum of the \KK pair.

The probability of producing an energy loss value $\varepsilon$ at a total momentum $p^\text{trk}$ for a given charged particle species $k$ 
($k$ can be \PGppm, \PKpm, proton \Pp, or antiproton \HepAntiParticle{\Pp}{}{}) can be approximated by
\begin{equation}
   \label{eq:dedxPDF}
   P_k(\varepsilon | p^\text{trk}) = \frac{1}{\sqrt{2\pi}\sigma_k} \exp\left(-\frac{( \varepsilon - \mu_k)^2}{2\sigma_k^2}\right),
\end{equation}
where $\varepsilon = \log(\dEdx)$, and $\mu_k$ and $\sigma_k$ are the corresponding values for the momentum dependent mean and width, respectively, of the $\varepsilon$ distribution for species $k$.
As the pair production of \Pp and \HepAntiParticle{\Pp}{}{} is strongly suppressed because of the lack of coupling to VM photoproductions in UPC events~\cite{Baltz:2007kq}, the selected two-track events mainly contain pions and kaons. To select high-quality \KK events, 
each track is required to satisfy $P_{\PK}(\varepsilon | p^\text{trk}) > 10 P_{\PGp}(\varepsilon | p^\text{trk})$.

The \KK pairs with an invariant mass within a range of $0.98 < \mKK < 1.15 \GeV$ are selected for the signal extraction of \PGf mesons.
These \KK pairs mainly include both $\PGf \to \KK$ decays and direct \KK pairs from photon-nuclear interactions. 
Direct pairs are produced when a photon fluctuates into a virtual \KK pair and elastically scatters off the target nucleus, forming a real \KK pair. 
These pairs contribute as a smooth continuum background in the \mKK distribution.
Contamination from misidentified \pipi pairs originating from coherently produced \PGr decays and direct \pipi pairs is estimated by assigning the \PK mass to the charged pions produced in Monte Carlo (MC) samples.
The pion pair contamination contributes only a negligible ${\sim} 0.01\%$ to the total observed yield in the signal mass range.

\begin{figure*}[hbt!]
   \centering   
   \includegraphics[keepaspectratio,width=0.49\textwidth]{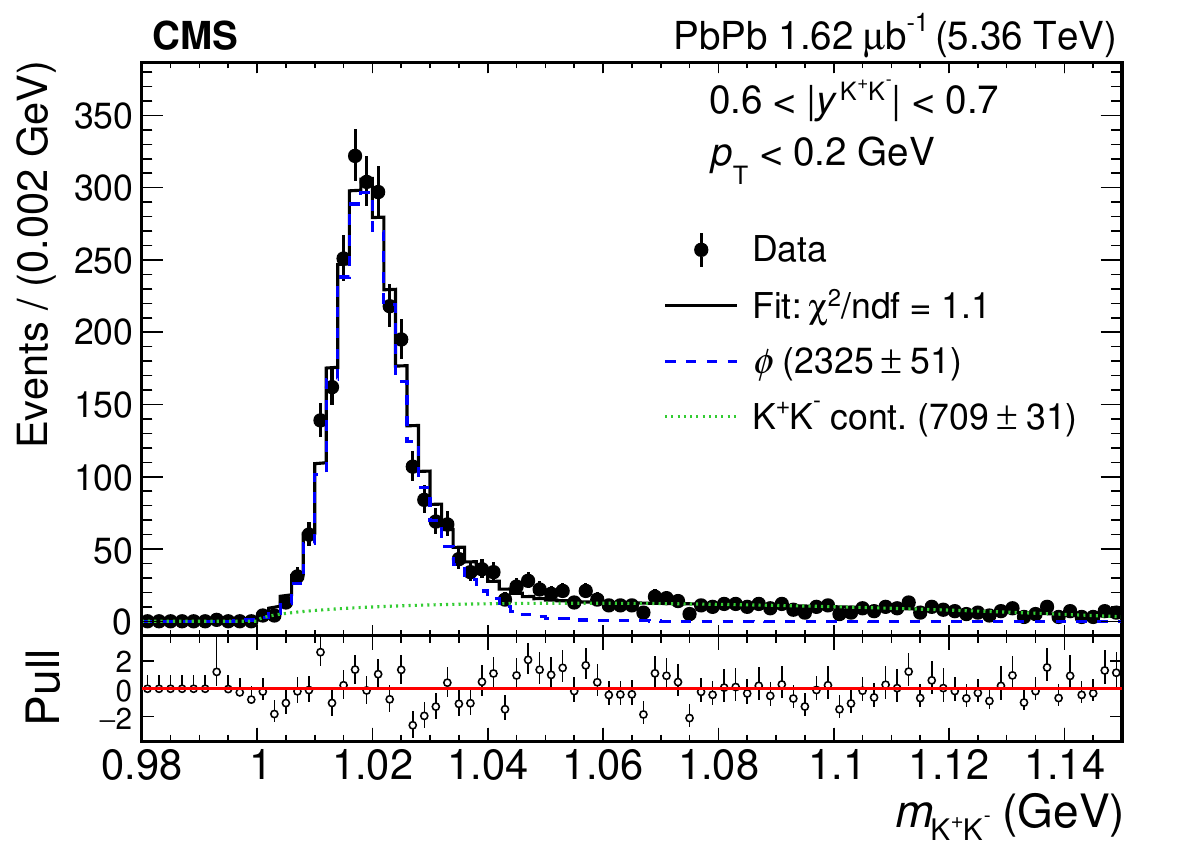}
   \includegraphics[keepaspectratio,width=0.49\textwidth]{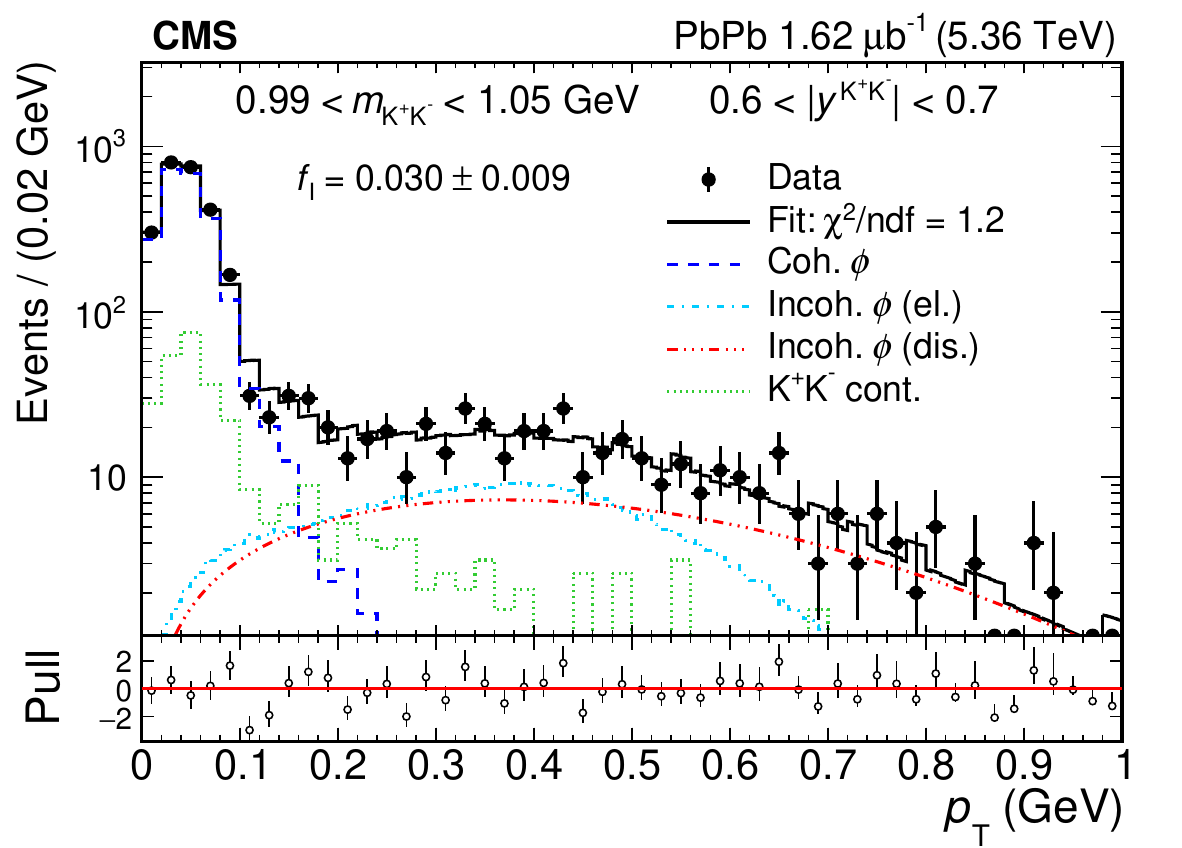}
   \caption{
      Examples of the invariant mass (\mKK) distribution in the coherent process dominant region of pair $\pt < 0.2 \GeV$ (left) and the pair \pt distribution in the \PGf meson mass window $0.99 < \mKK < 1.05 \GeV$ (right).
      The curves show results of the template fit, including the coherent (\text{Coh.}) \PGf, incoherent (\text{Incoh.}) \PGf for both elastic (el.) and dissociative (dis.) processes, and the direct \KK continuum (cont.).
      The error bars represent statistical uncertainties ($\sigma_{\text{Data}}$).
      The bottom panels show fit pulls, $(\text{Data} - \text{Fit}) / \sigma_{\text{Data}}$, with red lines at zero.
   }
   \label{fig:SignalsFitting}
\end{figure*}

The \PGf meson raw yield (\Ncoh) is extracted via a $\chi^2$ template fit to the \mKK distribution in the low \KK transverse momentum region $\pt{<}0.2\GeV$, where the coherent process dominates, as illustrated in Fig.~\ref{fig:SignalsFitting} (left).
The signal template shape of coherent \PGf meson production is obtained from the \STARlight (v3.30) event generator~\cite{Klein:2016yzr}, together with a full CMS detector simulation, as implemented with \GEANTfour~\cite{Agostinelli:2002hh}. 
Similar to previous studies~\cite{ZEUS:1996esk, CLAS:2007xhu, CMS:2016zzh}, the continuum background of direct \KK pairs is described by an empirical function $f_{\text{bkg}}(m) = A (m - 2 m_{\PK})^{1/2} + B (m - 2 m_{\PK})^{3/2}$, where $A$ and $B$ are free parameters. This function accounts for the experimental threshold effects near twice the kaon mass. 
A recent ALICE study~\cite{ALICE:2023kgv} reported a hint of enhanced direct \KK yield in the mass range $1.1<\mKK<1.4\GeV$ with a significance of $2.1\sigma$. This enhancement is attributed to potential interference effects between $\PGf \to \KK$ decay and direct \KK production. However, the absence of a direct \PGf meson measurement prevents a definitive conclusion regarding this explanation. A theoretical study~\cite{Ryskin:1997rh} suggests that such interference effects are negligible near the \PGf meson peak and become significant only at higher masses ($\mKK>1.2\GeV$). In this analysis, the $\chi^2/\text{ndf}$ (where ndf is the number of degrees of freedom) value and the pull distribution, as shown in the lower panel of Fig.~\ref{fig:SignalsFitting} (left), indicate that the fit without incorporating interference effects provides an accurate description of the data. This observation is consistent with the expectation of Ref.~\cite{Ryskin:1997rh}.

The extracted \Ncoh value includes contributions from both coherent and incoherent photoproduction. 
The fraction of the incoherent contribution (\fI) is determined by fits to the pair \pt distributions within the \PGf mass window.
Incoherent processes can be either elastic or dissociative, depending on whether the interacting nucleon remains intact or dissociates after the interaction, respectively.
The template \pt shapes of coherent and elastic incoherent \PGf meson photoproduction processes are from the \STARlight simulations.
The \pt distribution of the dissociative incoherent process is described by the function $\ddinline{N}{\pt} \approx \pt e^{b_{\text{pd}} \pt^{2}}$, as employed by the ZEUS experiment~\cite{ZEUS:1999ptu}, where $b_{\text{pd}}$ is a free parameter in the fit.
The \pt distribution of the \KK continuum is obtained from the higher-sideband region of the \PGf signal peak ($1.05 < \mKK < 1.15 \GeV$), 
while its normalization factor is fixed to the yield within the \PGf mass window determined in the full \pt range up to 1\GeV.
The default configuration of \STARlight does not perfectly simulate the \pt spectra of coherent VM photoproductions observed in recent measurements~\cite{CMS:2023snh, STAR:2022wfe, Acharya:2019vlb, ALICE:2021gpt, ALICE:2015nbw, ALICE:2021jnv}. 
In Ref.~\cite{CMS:2023snh}, it was pointed out that this issue can be overcome by enlarging the nuclear radius parameter. 
Changing the Pb radius parameter from 6.67 to 8.67\unit{fm} allows \STARlight to match the measured \PGf \pt spectra and, consequently, the larger radius value is used here.
In the fit, the normalization factors are free for the coherent, elastic incoherent and dissociative incoherent \PGf meson photoproduction.
The incoherent contamination fraction can be obtained as $\fI = [N_{\text{incoh}}^{\text{el}} + N_{\text{incoh}}^{\text{dis}}]/N_{\text{coh}}$ in the low-\pt ($<0.2\GeV$) region.
A representative fit to the \pt spectrum of the \KK pairs within the \PGf meson mass window is shown in Fig.~\ref{fig:SignalsFitting} (right).
The extracted \fI values in different rapidity intervals are found to be in the range 3--5\%.

The differential cross section of coherent  \PGf meson photoproduction is determined using the expression
\begin{equation}
   \DSigDyPhi = \frac{\Ncoh}{(1 + \fI) (\mathcal{A} \epsilon)_\PGf \mathcal{B}_{\PGf \to \KK} \lumi_{\text{int}} \Delta y},
\end{equation}
where \Ncoh is the \PGf meson yield extracted within $\pt<0.2 \GeV$, \fI is the incoherent fraction, 
$(\mathcal{A} \epsilon)_\PGf$ is the product of detector acceptance and efficiency for reconstructing the coherent \PGf 
meson photoproduction, $\mathcal{B}_{\PGf \to \KK} = (49.1 \pm 0.5)\%$ is the world-average branching 
fraction~\cite{ParticleDataGroup:2024cfk}, $\Delta y$ is the width of the rapidity interval, 
and $\lumi_{\text{int}}$ is the integrated luminosity of the data sample. 
The $(\mathcal{A} \epsilon)_\PGf$ factor is determined from simulation,
with values between 12--36\% for the \PGf meson rapidity range of $0.3 < \abs{y} < 1.0$. 
The dominant efficiency loss arises from the requirement on the minimum number of hits per track.
This condition is less often satisfied for tracks at midrapidity, where large curvature in the strong magnetic field reduces the number of tracker hits.
Additional corrections using the ``tag-and-probe'' (TnP) method~\cite{Chatrchyan:2012xi,CMS:2011aa} 
are applied to account for data-to-simulation discrepancies in the description of the online trigger 
and kaon reconstruction efficiencies. 
In this approach, one kaon track from $\PGf \to \KK$ decays is tightly selected (``tag''), while the second (``probe'') 
is a loosely reconstructed pixel-seed track used to determine the probability of satisfying the online trigger, 
as well as the offline full reconstruction and PID criteria. 
The resulting scale factors are about 0.85 in the $\PGf$ meson rapidity range. 
The reported cross section does not include contributions from coincident ultraperipheral collisions; any residual effect ($<2$\%) is included in the systematic uncertainty.

Systematic uncertainties are evaluated by taking the maximum deviation from the nominal result for each given uncertainty source. 
The uncertainties are 6.4\% in the integrated luminosity and 1\% in the $\mathcal{B}_{\PGf \to \KK}$~\cite{ParticleDataGroup:2024cfk}. 
Uncertainties propagated from the TnP scaling factors are estimated to be 2--6\%.
The threshold energies in the HF calorimeter used to reject hadronic contamination are tightened on both sides, which leads to a 2--3\% uncertainty. 
The uncertainty from the raw signal yield extraction is evaluated by using the alternative background 
model employed in Ref.~\cite{CLAS:2007xhu} and by varying the range of the \mKK for the fit. 
Overall, the uncertainties from signal yield extraction are found to be 3--6\%.
The systematic uncertainty in the incoherent fraction is estimated by 
(i) using the coherent \PGf meson production template with the default nuclear parameters in \STARlight, 
(ii) varying the range of the mass sideband for the \KK continuum template, and
(iii) removing the elastic incoherent \PGf template from the fit for the incoherent fraction estimation. 
The uncertainty from the incoherent fraction is less than 1\%. 
Individual sources of systematic uncertainties are added in quadrature to obtain the total systematic uncertainty of 7--10\%.

\begin{figure}[hbt!]
   \centering
   \includegraphics[keepaspectratio,width=0.95\linewidth]{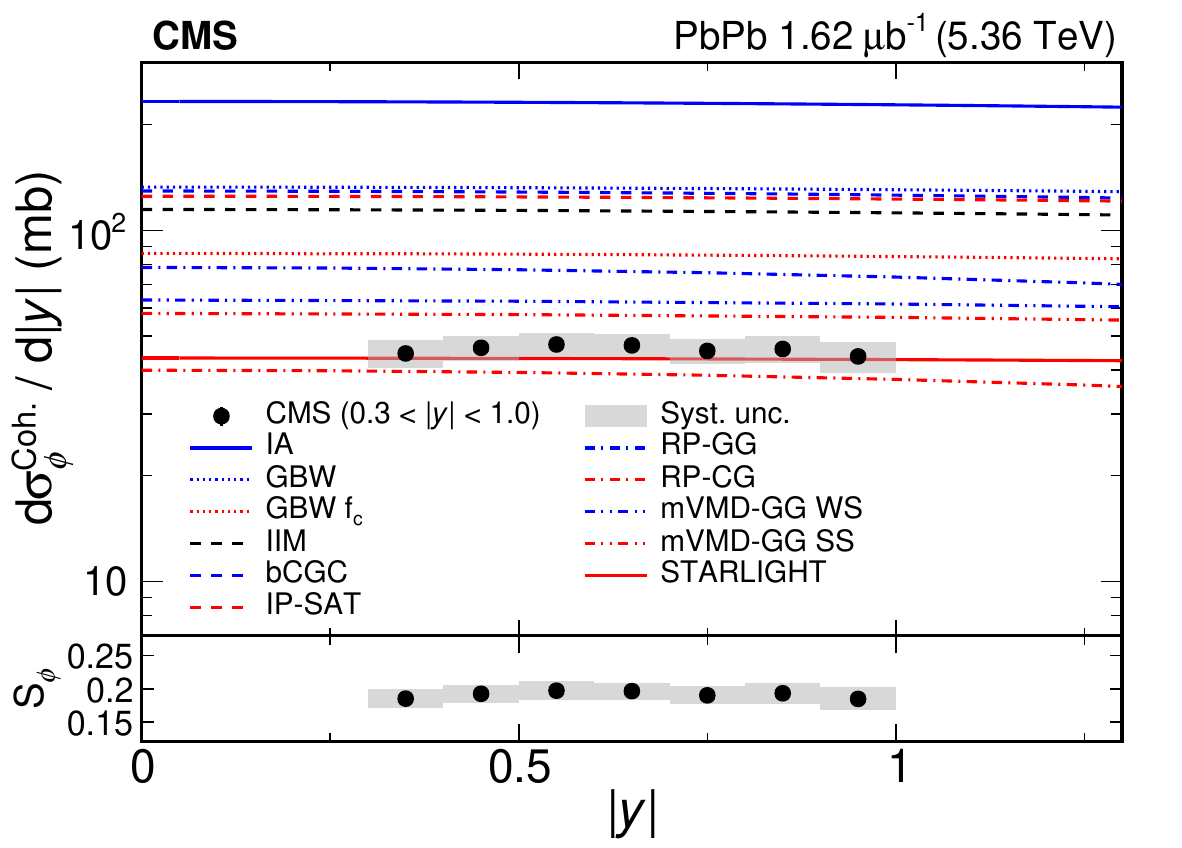}
   \caption{
The differential cross section of coherent $\PGf$ meson photoproduction as a function of rapidity in the range $0.3 < \abs{y} < 1.0$ (top panel).  
Vertical bars and shaded bands indicate statistical and systematic uncertainties, respectively.  
Theoretical predictions shown include the impulse approximation (IA)~\cite{Klein:2016yzr}, color dipole models (IIM~\cite{Iancu:2003ge}, bCGC~\cite{Kowalski:2006hc}, IP-SAT~\cite{Bartels:2002cj,Rezaeian:2012ji}, GBW and GBW $\text{f}_\text{c}$~\cite{Golec-Biernat:1998zce, Goncalves:2020cir}), reggeometric pomeron (RP) models with classical Glauber (CG) and Gribov--Glauber (GG) approaches~\cite{Jenkovszky:2022qnc, Jenkovszky:2023jrh},  
the modified VMD (mVMD) model with GG and strong or weak shadowing (SS or WS)~\cite{Guzey:2016piu}, and \STARlight with CG~\cite{Klein:2016yzr}.  
The bottom panel shows the data-to-IA ratio. 
      }
   \label{fig:DSigmaDy_vs_y}
\end{figure}

The differential cross section for coherent \PGf meson photoproduction over the rapidity range $0.3 < \abs{y} < 1.0$ is shown in Fig.~\ref{fig:DSigmaDy_vs_y}.
The characteristic $x$ values of the probed gluons in a given \abs{y} interval can be obtained using
\begin{equation}
   \label{eq:x}
   x = \frac{m_{\PGf}}{\sqrtsNN} \exp(\pm y),
\end{equation}
where the sign of $y$ corresponds to the rapidity sign of the photon-emitting nucleus in the laboratory frame~\cite{Jones:2013pga, Guzey:2013jaa, CMS:2023snh, CMS:2025oxg}. However, in symmetric UPCs, the photon emitter cannot be definitively identified~\cite{Guzey:2013jaa,Guzey:2016piu}. Consequently, the cross section values at each \abs{y} represent the average contributions from $+y$ and $-y$, without separating individual contributions. The $x$ values corresponding to the \abs{y} coverage are around $10^{-4}$, falling within the range explored in recent measurements~\cite{CMS:2023snh,ALICE:2023jgu}, where the coherent \PJGy meson photoproduction cross section exhibits strong nuclear suppression effects~\cite{Mantysaari:2024zxq}.

The data are compared to various theoretical models, all of which are constrained to photoproduction data off protons from HERA~\cite{ZEUS:1996esk}, either directly or via fitted parameterizations (as in the case of \STARlight).
A baseline model, referred to as the impulse approximation (IA), represents the direct sum of coherent contributions by treating the nucleus as a simple collection of free nucleons~\cite{Klein:1999qj,Guzey:2016piu}. 
To quantify the impact of nuclear modifications on coherent \PGf meson photoproduction, a nuclear suppression factor, \Sphi, is defined as the ratio of the measured cross section to the IA prediction~\cite{Guzey:2016piu}.
The extracted \Sphi values, as shown in the bottom panel of Fig.~\ref{fig:DSigmaDy_vs_y}, are found to be ${\sim} 0.18{-}0.20$ across the entire \abs{y} range. 
This observation provides strong evidence that coherent \PGf meson photoproduction undergoes significant nuclear suppression.
Moreover, the value of \Sphi is approximately half of \Sjpsi ($0.41 \pm 0.05$) within the $0<\abs{y}<0.8$ range, as reported by the ALICE experiment~\cite{ALICE:2021gpt} with \PbPb UPCs at $\sqrtsNN = 5.02\TeV$. The greater suppression of the \PGf meson confirms the mass dependence in coherent VM photoproductions~\cite{Forshaw:2004vv, Kowalski:2006hc, Frankfurt:2005mc}.

Theoretical calculations that include gluon-saturation effects at small $x$ are performed within the color glass condensate (CGC) framework~\cite{Gelis:2010nm} using the color-dipole formalism~\cite{Goncalves:2017wgg}.  Three representative approaches are compared with the data. 
The Iancu--Itakura--Munier (IIM) model~\cite{Iancu:2003ge} interpolates between the Balitsky--Fadin--Kuraev--Lipatov~\cite{Kuraev:1977fs,Balitsky:1978ic,Kovchegov:1999yj} and Balitsky--Kovchegov~\cite{Balitsky:1995ub,Kovchegov:1999yj,Kovchegov:1999ua} equations, the bCGC model~\cite{Kowalski:2006hc} adds impact-parameter dependence to the saturation scale, and the IP-SAT model~\cite{Bartels:2002cj,Rezaeian:2012ji} employs an eikonalized dipole amplitude based on a DGLAP-evolved gluon density~\cite{Gribov:1972ri,Altarelli:1977zs,Dokshitzer:1977sg}. 
All three models overpredict the measured coherent $\PGf$ photoproduction cross section off lead nuclei by a factor of 2.6--3.0.

Notably, a theoretical study~\cite{Goncalves:2020cir}, based on a framework similar to the bCGC model and known as the Golec--Biernat--Wusthoff (GBW) model, specifically investigates non-pQCD corrections to the photon wave function within the color dipole formalism. The predictions without and with these corrections, referred to as GBW and GBW $\text{f}_\text{c}$, respectively, are shown in Fig.~\ref{fig:DSigmaDy_vs_y}. The inclusion of non-pQCD corrections results in an additional reduction of approximately 40\% in the predicted cross sections, but the predictions still overestimate the data by a factor of about 2.

In contrast to gluon-saturation models, another class of calculations accounts for non-linear QCD effects at small~$x$ through \emph{nuclear shadowing}. 
These models generally use the vector meson dominance (VMD) picture~\cite{Bauer:1977iq} combined with either the classical Glauber (CG)~\cite{Gribov:1968jf} or the Gribov--Glauber (GG)~\cite{Glauber:1970jm} formalism to describe multiple scatterings of the virtual vector meson in the nucleus. 
While CG includes only elastic interactions, GG extends the treatment to inelastic channels. 
The reggeometric Pomeron (RP) model~\cite{Jenkovszky:2022qnc,Jenkovszky:2023jrh} combines VMD with both CG and GG schemes, with the latter predicting cross sections roughly twice as large. 
The CG-based results and the modified VMD (mVMD) approach with strong or weak shadowing scenarios~\cite{Guzey:2016piu} describe the data more closely, though they still overshoot by 25--35\%. 
Predictions from the \textsc{STARlight} model~\cite{Klein:2016yzr}, which employs VMD with the CG formalism, agree with the measurements within a few percent, despite its known limitations for coherent $\PGrzP{770}$ and \PJGy production~\cite{ALICE:2021gpt,ALICE:2020ugp,LHCb:2021bfl,CMS:2023snh,ALICE:2023jgu,STAR:2023nos}.

In summary, the first measurement of coherent \Pgf meson photoproduction off a heavy nucleus has been performed by the CMS experiment in ultraperipheral lead-lead collisions at a center-of-mass energy per nucleon pair of 5.36\TeV.
The production cross section is reported in the \PGf meson rapidity range $0.3 < \abs{y} < 1.0$, probing gluons within the lead nucleus having $x$ values around $10^{-4}$ at an intermediate energy scale set by the \PGf meson mass.
This places it in a unique regime between perturbative and nonperturbative quantum chromodynamics.
The cross section is found to be significantly suppressed by a factor of ${\sim} 5$ compared to an impulse approximation model that treats a nucleus as a collection of free nucleons.
Models that account for nuclear shadowing effects generally provide better predictions of the data compared to those that incorporate gluon saturation effects.
The measurement of coherent \PGf meson photoproduction off a heavy nucleus establishes a powerful new tool for exploring nuclear effects and nuclear gluonic structure in the small-$x$ regime.

\begin{acknowledgments}
   We congratulate our colleagues in the CERN accelerator departments for the excellent performance of the LHC and thank the technical and administrative staffs at CERN and at other CMS institutes for their contributions to the success of the CMS effort. In addition, we gratefully acknowledge the computing centers and personnel of the Worldwide LHC Computing Grid and other centers for delivering so effectively the computing infrastructure essential to our analyses. Finally, we acknowledge the enduring support for the construction and operation of the LHC, the CMS detector, and the supporting computing infrastructure provided by the following funding agencies: SC (Armenia), BMBWF and FWF (Austria); FNRS and FWO (Belgium); CNPq, CAPES, FAPERJ, FAPERGS, and FAPESP (Brazil); MES and BNSF (Bulgaria); CERN; CAS, MoST, and NSFC (China); MINCIENCIAS (Colombia); MSES and CSF (Croatia); RIF (Cyprus); SENESCYT (Ecuador); ERC PRG, RVTT3 and MoER TK202 (Estonia); Academy of Finland, MEC, and HIP (Finland); CEA and CNRS/IN2P3 (France); SRNSF (Georgia); BMBF, DFG, and HGF (Germany); GSRI (Greece); NKFIH (Hungary); DAE and DST (India); IPM (Iran); SFI (Ireland); INFN (Italy); MSIP and NRF (Republic of Korea); MES (Latvia); LMTLT (Lithuania); MOE and UM (Malaysia); BUAP, CINVESTAV, CONACYT, LNS, SEP, and UASLP-FAI (Mexico); MOS (Montenegro); MBIE (New Zealand); PAEC (Pakistan); MES and NSC (Poland); FCT (Portugal); MESTD (Serbia); MICIU/AEI and PCTI (Spain); MOSTR (Sri Lanka); Swiss Funding Agencies (Switzerland); MST (Taipei); MHESI and NSTDA (Thailand); TUBITAK and TENMAK (Turkey); NASU (Ukraine); STFC (United Kingdom); DOE and NSF (USA).
\end{acknowledgments}

\bibliography{auto_generated}

\cleardoublepage \appendix\section{The CMS Collaboration \label{app:collab}}\begin{sloppypar}\hyphenpenalty=5000\widowpenalty=500\clubpenalty=5000\cmsinstitute{Yerevan Physics Institute, Yerevan, Armenia}
{\tolerance=6000
V.~Chekhovsky, A.~Hayrapetyan, V.~Makarenko\cmsorcid{0000-0002-8406-8605}, A.~Tumasyan\cmsAuthorMark{1}\cmsorcid{0009-0000-0684-6742}
\par}
\cmsinstitute{Institut f\"{u}r Hochenergiephysik, Vienna, Austria}
{\tolerance=6000
W.~Adam\cmsorcid{0000-0001-9099-4341}, J.W.~Andrejkovic, L.~Benato\cmsorcid{0000-0001-5135-7489}, T.~Bergauer\cmsorcid{0000-0002-5786-0293}, K.~Damanakis\cmsorcid{0000-0001-5389-2872}, M.~Dragicevic\cmsorcid{0000-0003-1967-6783}, C.~Giordano, P.S.~Hussain\cmsorcid{0000-0002-4825-5278}, M.~Jeitler\cmsAuthorMark{2}\cmsorcid{0000-0002-5141-9560}, N.~Krammer\cmsorcid{0000-0002-0548-0985}, A.~Li\cmsorcid{0000-0002-4547-116X}, D.~Liko\cmsorcid{0000-0002-3380-473X}, I.~Mikulec\cmsorcid{0000-0003-0385-2746}, J.~Schieck\cmsAuthorMark{2}\cmsorcid{0000-0002-1058-8093}, R.~Sch\"{o}fbeck\cmsAuthorMark{2}\cmsorcid{0000-0002-2332-8784}, D.~Schwarz\cmsorcid{0000-0002-3821-7331}, M.~Sonawane\cmsorcid{0000-0003-0510-7010}, W.~Waltenberger\cmsorcid{0000-0002-6215-7228}, C.-E.~Wulz\cmsAuthorMark{2}\cmsorcid{0000-0001-9226-5812}
% ...
\par}
\vskip\cmsinstskip
\dag:~Deceased\\
$^{1}$Also at Yerevan State University, Yerevan, Armenia\\
$^{2}$Also at TU Wien, Vienna, Austria\\
% ...
\end{sloppypar}
%%% END EDITABLE REGION %%%
% skeleton_end
\end{document}